\documentclass[%
reprint,
 amsmath,amssymb,
 aps, physrev,
]{revtex4-2}
\usepackage{amsmath,amsfonts,amssymb,graphicx,slashed}
\usepackage{dcolumn}
\usepackage{physics}
\usepackage[utf8]{inputenc}
\usepackage{geometry}
\usepackage{xcolor}
\usepackage{appendix}
\usepackage{bm}
\usepackage{xcolor}
\usepackage{amsmath}
\usepackage{amssymb}
\begin{document}

\title{Interacting Dark Sector field theory with phantom crossing}
\author{E. Abdalla$^{1,2}$}
 \email{eabdalla@usp.br}
\author{O. P. F. Piedra$^{1}$}
 \email{opavelfp2006@gmail.com}
\author{A. A. Escobal$^{3,4}$}
 \email{anderson@ustc.edu.cn}
\author{A. M. Vicente$^5$}
\author{F. B. Abdalla$^{3,4}$}
\author{B. Wang$^{6,7}$}
\author{A. Marins$^{3,4}$}
\affiliation{$^{1}$Department of Physics, Center for Exact and Natural Sciences, Federal University of Paraíba, 58059-970, João Pessoa, Brazil\\
$^2$ Secretary of Science and Technology, Government of Paraíba, Brazil\\
$^{3}$ Department of Astronomy, University of Science and Technology of China, Hefei, Anhui 230026, China.\\
$^4$ School of Astronomy and Space Science, University of Science and Technology of China, Hefei, Anhui 230026, China. \\
$^5$Department of Physics, University of S\~ao Paulo, 05508-900, S\~ao Paulo, SP, Brazil.\\
$^6$ Center for Gravitation and Cosmology, Yangzhou University, Yangzhou 224009, China \\
$^7$School of Aeronautics and Astronautics, Shanghai Jiao Tong University, Shanghai 200240, China\\ }
\date{\today}

\begin{abstract}
Recent results from the Dark Energy Spectroscopic Instrument (DESI) provide evidence for a dynamical dark-energy component, whose equation of state appears to have recently crossed the phantom divide. In this Letter, we present an interacting dark-energy model, grounded in field theory, that naturally accommodates such a double crossing. In our framework, fermionic dark matter is coupled via a Yukawa interaction to a tachyonic scalar field governed by Born–Infeld dynamics. The phantom crossing arises at the level of the effective dark-energy equation of state, while the underlying scalar-field dynamics remains nonphantom and well bounded. We confront our model with data including BAO from the DESI (DR2) survey, CMB distance priors from Planck 2018, and the latest Type Ia supernovae compilations, obtaining robust constraints across the different data combinations and reconstructing a recent double crossing of the phantom divide. Furthermore, under naturalness assumptions, the model expects an ultralight fermionic dark matter mass of order $1.9\times10^{-3}\,\mathrm{eV}$, suggesting a possible connection with new light particles in the dark sector and motivating future tests with cosmological perturbations. 

\end{abstract}

\maketitle
\textit{Introduction}---The origin of late-time cosmic acceleration remains a significant challenge in modern cosmology. The standard $\Lambda$CDM model describes dark energy as a cosmological constant. In view of some theoretical questions, there are several attempts to describe Dark Energy as a dynamical component of the Universe \cite{Wang:2016lxa,Wang:2024vmw}. Furthermos, there are indications from observations that dynamical dark energy  equation of state might cross the phantom divide \cite{DESI2025}. Addressing this question has long motivated the development of several classes of physical models, including quintom scenarios, effective descriptions with additional degrees of freedom, and interacting dark-sector models \cite{FENG200535,GUO2005177,PhysRevD.71.047301,Wang:2005jx,CAI20101}. The recent measurements from the Dark Energy Spectroscopic Instrument (DESI) and the cosmological analysis by the DESI collaboration \cite{DESI2025} have enhanced interest in dynamical Dark Energy models, as they indicate that the inferred evolution of the Dark Energy equation of state can be compatible with a transition across the phantom divide \cite{thanankullaphong2026,10.1093/mnras/staf1927,10.1093/nsr/nwag115}.

Constructing a minimal and fully consistent theoretical realization of such a crossing, however, remains an open problem. A fundamental phantom field typically involves negative kinetic energy and associated stability problems, while a sustained phantom regime can lead to future singularities \cite{CALDWELL200223,PhysRevD.68.023509}. Avoiding these pathologies has motivated alternative frameworks, including D3-brane and braneworld constructions \cite{PhysRevD.72.064017,Mishra:2025goj}, modified-gravity interpretations \cite{PhysRevD.110.123524}, and effective mechanisms that allow a regular crossing of $w=-1$ \cite{PhysRevD.78.087303}.

Among the proposed mechanisms to avoid a fundamental phantom degree of freedom, which is responsible for crossing the phantom divide, interacting dark-sector models constitute well-motivated alternatives \cite{Wang:2005jx}. In these scenarios, the crossing can arise from the modification of the energy balance between dark matter and dark energy, and not from a phantom kinetic term. However, in many existing constructions, the energy transfer is introduced phenomenologically through a prescribed interaction term $Q$ in the coupled cosmological balance equations for the dark components (see \cite{VANDERWESTHUIZEN2025102119,Wang:2016lxa,Escobal:2026zxb,Wang:2024vmw,Hoerning:2023hks} and references therein). This leaves open the question of whether the crossing of the phantom divide can arise from a microscopic, dynamical, and self-consistent dark-sector model. In fact, consistent phenomenological model that incorporate phantom crossing have already being discussed in the literature \cite{Wang:2005ph,Wang:2005jx}.

In this Letter, we extend beyond a mere phenomenological description to present a stable and minimal realization of phantom crossing through fundamental dark-sector fields. A field theory approach has been similarly advocated previously \cite{Costa:2014pba,Micheletti:2009pk}. In this work, we progress further by characterizing Dark Energy as a tachyonic scalar field with Born--Infeld (BI) dynamics ($\varphi$), while modeling dark matter as a massive Dirac fermion ($\Psi$). The dark sector fields are coupled through a Yukawa-type interaction \cite{Micheletti:2009pk,Costa:2014pba}. Such a microscopic interaction may induce an effective double crossing of the phantom divide. When the model is confronted with recent cosmological data sets, combining BAO measurements from DESI with CMB distance priors and Type Ia supernova observations, the posterior reconstruction of the effective equation of state of $\varphi$ indicates that the double crossing persists for two standard deviation. 

\textit{Theory}---The proposed Lagrangian density describing the dynamics of the dark sector is \cite{Micheletti:2009pk}\footnote{We adopt the metric signature $(+,-,-,-)$, and define the covariant derivative as $\nabla_\mu \Psi = \partial_\mu \Psi + \frac{1}{4}\omega_{\mu AB}\gamma^{AB}\Psi$, where $\omega_{\mu AB}$ denotes the spin connection, where $\mu=0$ corresponds to the cosmic time $t$. We also denote the derivative of a quantity $f$ with respect to $t$ as $\dot{f}$, and use units $\hbar=c=1$.}
\begin{align}
\mathcal{L}_M &= -V(\varphi)\sqrt{1 - \alpha \partial^\mu \varphi \partial_\mu \varphi}
+\frac{i}{2}\left(\bar{\Psi} \overleftrightarrow{\nabla}_\mu \hat{\gamma}^\mu \Psi\right)\nonumber\\\
&\phantom{=}- (M - \beta \varphi)\bar{\Psi}\Psi\quad .
\label{eq:lagrangianinteraction}
\end{align}
where the tachyonic scalar field $\varphi$ with BI–type dynamics and potential $V(\phi)$ describes dark energy, and the massive Dirac fermion field $\Psi$ models dark matter. The constant $M$ denotes the bare fermion mass, and the Yukawa-type interaction with the scalar field generates the effective fermion mass $M_{\rm eff}(\varphi)=M-\beta\varphi.$
Here, the dimensionless parameter $\beta$ controls the strength of the scalar-fermion coupling, whereas the parameter $\alpha>0$, with dimensions of mass$^{-4}$, sets the kinetic scale of the BI term.

In a homogeneous FLRW spacetime, the solution of the Dirac equation derived from the above Lagrangian density gives, for the fermion bilinear $n_{\Psi}(t) = \bar{\Psi}(t)\Psi(t)$, the redshift-dependence characteristic of pressureless matter, $n_{\Psi}(z) = n_{\Psi,0}(1+z)^3$.
For the scalar field, we find the equation of motion given by  
\begin{equation}
\frac{\ddot{\varphi}}{1 - \alpha \dot{\varphi}^2}=
-\frac{1}{\alpha}\frac{d\ln V(\varphi)}{d\varphi}-
3H\dot{\varphi}
+\frac{\beta n_\Psi}{\alpha V}\sqrt{1 - \alpha \dot{\varphi}^2}.
\label{eq:ddot_phi_from_compact}
\end{equation}

From the energy-momentum tensor derived from the above Lagrangian density, one obtains, in a FLRW geometry, the corresponding energy densities and pressures. For the fermionic field, these are given by $\rho_\Psi = \left(M-\beta\varphi\right)\bar{\Psi}\Psi$ and $p_\Psi = 0$, while for the scalar field one finds $\rho_\varphi = V(\varphi)/\sqrt{1-\alpha\dot{\varphi}^{\,2}}$ and $p_\varphi = -V(\varphi)\sqrt{1-\alpha\dot{\varphi}^{\,2}}$. Therefore, the tachyonic dark-energy component is dynamical, with equation of state (EoS) parameter $w_\varphi = \alpha\dot{\varphi}^{\,2}-1$. In this case, the scalar field alone cannot cross the phantom divide, as a consequence of the BI kinematical restriction $\alpha\dot{\varphi}^{\,2}<1$, as can be seen in Eq. \ref{eq:ddot_phi_from_compact}.

The Friedmann equations lead to the coupled continuity relations for the scalar and fermionic fields,
$\dot{\rho}_\varphi = Q - 3H\alpha\dot{\varphi}^{\,2}\rho_\varphi$ and
$\dot{\rho}_\Psi = -Q - 3H\rho_\Psi$, respectively. The Hubble expansion can then be written in terms of the components of the Universe as
\begin{align}
3M_{\rm Pl}^2H^2
&=
\rho_{r,0}(1+z)^4+\rho_{b,0}(1+z)^3\nonumber \\
&\phantom{=}+(M-\beta\varphi)n_{\Psi}+\frac{V(\varphi)}{\sqrt{1-\alpha\dot\varphi^2}},\label{eq:Friedmann_rest_original}
\end{align}
where $M_{\mathrm{Pl}}=1/\sqrt{8 \pi G}$ is the reduced Planck mass. The interaction kernel $Q = \beta \dot{\varphi} n_\Psi$ quantifies the rate of energy transfer between the two components. For $\beta > 0$ and $n_\Psi > 0$, an increasing scalar field ($\dot{\varphi} > 0$) implies $Q > 0$, corresponding to a flow of energy from the fermionic sector to the scalar field. At this point, it is clear that the transition from the phenomenological models \cite{Wang:2016lxa,Hoerning:2023hks} and field theory models \cite{Costa:2014pba} has the advantage of describing the EoS of Dark Energy in terms of fundamental fields. 

The interaction modifies the cosmological dynamics in a nontrivial way. To capture this effect at the background level, we define an effective EoS for the scalar component through the standard continuity equation, $\dot{\rho}_\varphi + 3H\bigl(1 + w_{\mathrm{eff}}^{\varphi}\bigr)\rho_\varphi = 0$, which describes the macroscopic evolution of the scalar-field energy density as if it were separately conserved. Then, we obtain the effective EoS
\begin{equation}
w^{\mathrm{eff}}_{\varphi}
=
w_\varphi - \frac{\beta \dot{\varphi} n_\Psi}{3H \rho_\varphi}
= -1+ \left( \alpha \dot{\varphi}^2 - \frac{\beta \dot{\varphi} n_\Psi}{3H \rho_\varphi}
\right).
\label{eq:effective_EoS}
\end{equation}
where the second term inside the parentheses on the right-hand side is the contribution induced by the scalar--fermion interaction, and represents the effect of energy exchange between the dark-sector components, normalized by the Hubble dilution scale $3H\rho_\varphi$.

To confront the above field-theoretic model with background cosmological data, it is useful to rewrite the dark sector dimensional parameters in terms of quantities adapted to the hierarchy between gravity and late-time cosmic expansion. The relevant energy scales are the reduced Planck mass, $M_{\rm Pl}=2.435\times10^{18}\,\mathrm{GeV}$, and the energy scale associated with the present Hubble rate, $M_H\equiv \hbar H_0=1.5\times10^{-33}\,\mathrm{eV}$. It is therefore convenient to absorb the dimensions of the model parameters into this hierarchy. Hence, we can rescale the scalar field through these two mass scales as $\alpha=1/(M_HM_{\rm Pl})^2$,  and parametrize the fermion mass and the scalar--fermion coupling as $M=\widetilde{M}\sqrt{M_HM_{\rm Pl}}$ and $\beta=\tilde{\beta}\sqrt{M_H/M_{\rm Pl}}$, where the tilde quantities are dimensionless parameters entering the Bayesian forecast analysis. This parametrization factors out the large hierarchy between the cosmological and Planck scales, leaving dimensionless combinations to be constrained by the background evolution.

\textit{Numerical}--- For the numerical purposes, to investigate the cosmological dynamics, we adopted an exponential potential $V(\varphi)=V_0e^{-\eta\varphi}$ for the scalar sector, and rewrote the evolution equation as two first-order differential equations in the redshift variable $z$, for the dimensionless unknown functions ($Y(z),
u(z)$), defined as $Y=(\varphi-\varphi_0)/M_{\rm Pl}$ and $ u=\operatorname{arctanh}\left(\dot{\varphi}/ H_0M_{\rm Pl}\right)$. The variable $u$ is introduced to ensure robust numerical evolution near the BI velocity bound. Indeed, the factors $\sqrt{1-\alpha\dot{\varphi}^{2}}$ appearing in Eqs.~\eqref{eq:Friedmann_rest_original} and \eqref{eq:ddot_phi_from_compact} vanish when the field approaches its terminal velocity, $|\dot{\varphi}|= H_0M_{\rm Pl}$, producing numerical instabilities. This is resolved by the choice of the variable $u$ above, which maps the physical allowed interval for the velocity into the full real line, $-\infty<u<\infty$.

The cosmological evolution equations can be written in a convenient closed form in terms of the scalar field as
\begin{align}
\frac{dY}{dz}
&=
-\frac{\tanh u}{(1+z)E}
\label{eq:system_Yu_1},
\\[6pt]
\frac{du}{dz} &= \frac{1}{(1+z)E} \left[ -\tilde\eta + 3E\tanh u\right.\nonumber \\
&\phantom{=}\left. - \frac{\zeta_{\rm eff}\Omega_{\Psi,0}e^{\tilde\eta Y}(1+z)^3}{\Omega_{\varphi,0}} \frac{\operatorname{sech}u}{\operatorname{sech}u_0}
\right],
\end{align}
Here, $\tilde{\eta}=\eta M_{\rm Pl}$, while $\zeta_{\rm eff}$ is an effective reparametrization that encodes the interaction strength and fermionic parameters, and it is defined as $\zeta_{\rm eff} = \zeta/(1-\zeta\varphi_0/M_{\rm Pl})$,
with $\zeta\equiv \tilde{\beta}/\tilde{M}$. The dimensionless Hubble function, $E(z)=H(z)/H_0$, is then expressed by
\begin{align}
E^2(z) &= \Omega_{r,0}(1+z)^4 + \Omega_{\varphi,0}e^{-\tilde\eta Y}\frac{\cosh u}{\cosh u_0}
\nonumber \\
&\quad+ \left[\Omega_{b,0}+\Omega_{\Psi,0}(1-\zeta_{\rm eff}Y)\right](1+z)^3.
\label{eq:Ez_reparametrized}
\end{align}

At the level of the homogeneous expansion, the dynamics is not sensitive to $M$, $\beta$, and the field normalization associated with $\alpha$ as independent microscopic quantities. Instead, as exposed in Eq. \ref{eq:Ez_reparametrized}, it depends on $\zeta_{\rm eff}$, which controls both the fermionic contribution to $E^2(z)$ and the interaction term in the effective scalar EoS. Therefore, background observables constrain this combination of effective interactions, rather than $M$, $\beta$, and $\alpha$ separately.

The resulting system is then integrated numerically, directly yielding $Y(z),
u(z)$ and $E(z)$.
The interaction kernel $Q$ now becomes
\begin{align}
Q
=
3M_{\mathrm{Pl}}^2 H_0^2 \zeta_{\mathrm{eff}}\Omega_{\Psi,0}(1+z)^3\tanh u
\label{eq:Qphi_final_Yu} \;,
\end{align}
and the effective EoS becomes
\begin{align}
w_{\varphi}^{\mathrm{eff}}(z)
&=-
\frac{e^{\tilde\eta Y(z)}\Omega_{\Psi,0}(1+z)^3\zeta_{\mathrm{eff}}\tanh u(z)\operatorname{sech}u(z)}
{3E(z)\Omega_{\varphi,0}\operatorname{sech}u_0}\nonumber
\\
&\phantom{=}
-\operatorname{sech}^2u(z),
\label{eq:eff_EOS_Yu}
\end{align}
where $w_{\varphi}=-\operatorname{sech}^2u(z)$ is the intrinsic tachyonic EoS.

\textit{Results}---In order to establish the viability of the proposed cosmological scenario, we must derive stringent constraints on its parameter space through a direct comparison with observational data. We perform a Bayesian parameter estimation utilizing the \textit{dynesty} package \cite{2020MNRAS.493.3132S,sergey_koposov_2025_17268284}, which employs a nested sampling algorithm. This approach is particularly advantageous for our model, as it is inherently well-suited for exploring multimodal posterior distributions—a feature that can pose significant challenges for standard Markov Chain Monte Carlo (MCMC) techniques.

The late-time expansion history is primarily anchored by the most recent Baryon Acoustic Oscillation (BAO) measurements from the DESI collaboration Data Release 2 (DR2) \cite{DESI:2025zgx}. Spanning a redshift interval of $0.295 < z < 2.33$, this dataset supplies critical geometrical information via the dimensionless ratios between cosmological $D_{M}/r_{s,\mathrm{drag}}$, $D_{H}/r_{s,\mathrm{drag}}$, and $D_{V}/r_{s,\mathrm{drag}}$ \footnote{Here, $D_{M}$ denotes the transverse comoving distance, $D_{H} \equiv c/H(z)$ is the Hubble distance, $D_{V}$ is the volume-averaged (isotropic) BAO distance, and $r_{s,\mathrm{drag}}$ is the comoving sound horizon at the drag epoch. The drag-epoch redshift $z_{\mathrm{drag}}$ is computed using the approximation described in \cite{Hu:1995en}.}. We complement these measurements with luminosity distance data from three state-of-the-art Type Ia Supernovae (SNe Ia) catalogs. The first is the Pantheon+ (PP) compilation \cite{Brout:2022vxf}, containing 1701 light curves from 1550 spectroscopically confirmed SNe Ia across $0.001 < z < 2.26$. The second is the Dark Energy Survey Year 5 (DESY5) sample \cite{DES:2024jxu}, which contributes 1829 photometrically classified events within $0.10 < z < 1.13$. The third catalog is Union3 \cite{Rubin:2023jdq}, offering 2087 cosmologically useful SNe Ia from various datasets spanning $0.001 < z < 2.26$. Because these compilations share partially overlapping observations, we treat each SNe Ia dataset independently throughout our analysis to prevent statistical redundancies. Furthermore, to effectively break parameter degeneracies, we incorporate the acoustic angular scale, $\ell_A$, extracted from the Planck 2018   Cosmic Microwave Background (CMB) \cite{Planck2018} distance priors \cite{Chen:2018dbv}. 

Regarding the statistical configuration, broad, flat prior distributions are imposed on all primary variables to ensure that the final parameter space is observationally driven rather than prior-dominated. The adopted uniform priors for the free parameters are summarized in Table~\ref{tab:Prior}.
\begin{table}[h]
\centering

\renewcommand{\arraystretch}{1.2}
\begin{tabular}{c c}
\hline\hline
Parameter & Prior Range \\
\hline
$H_0$ [$\mathrm{km\,s^{-1}\,Mpc^{-1}}$] & $[50.0, 90.0]$ \\
$\Omega_{\Psi,0}$ & $[0.001, 0.8]$ \\
$\Omega_{b,0}$ & $[0.001, 0.09]$ \\
$\zeta_{\mathrm{eff}}$ & $[-45.0, 30.0]$ \\
$\tilde{\eta}$ & $[0.0, 9.0]$ \\
${\dot{\varphi}_0}/({H_0M_{\text{Pl}}})$ & $[-1.0, 1.0]$ \\
\hline\hline
\end{tabular}
\caption{Uniform priors on the free parameters of the model.}
\label{tab:Prior}
\end{table}

\begin{figure}
    \centering \includegraphics[width=0.5\textwidth]{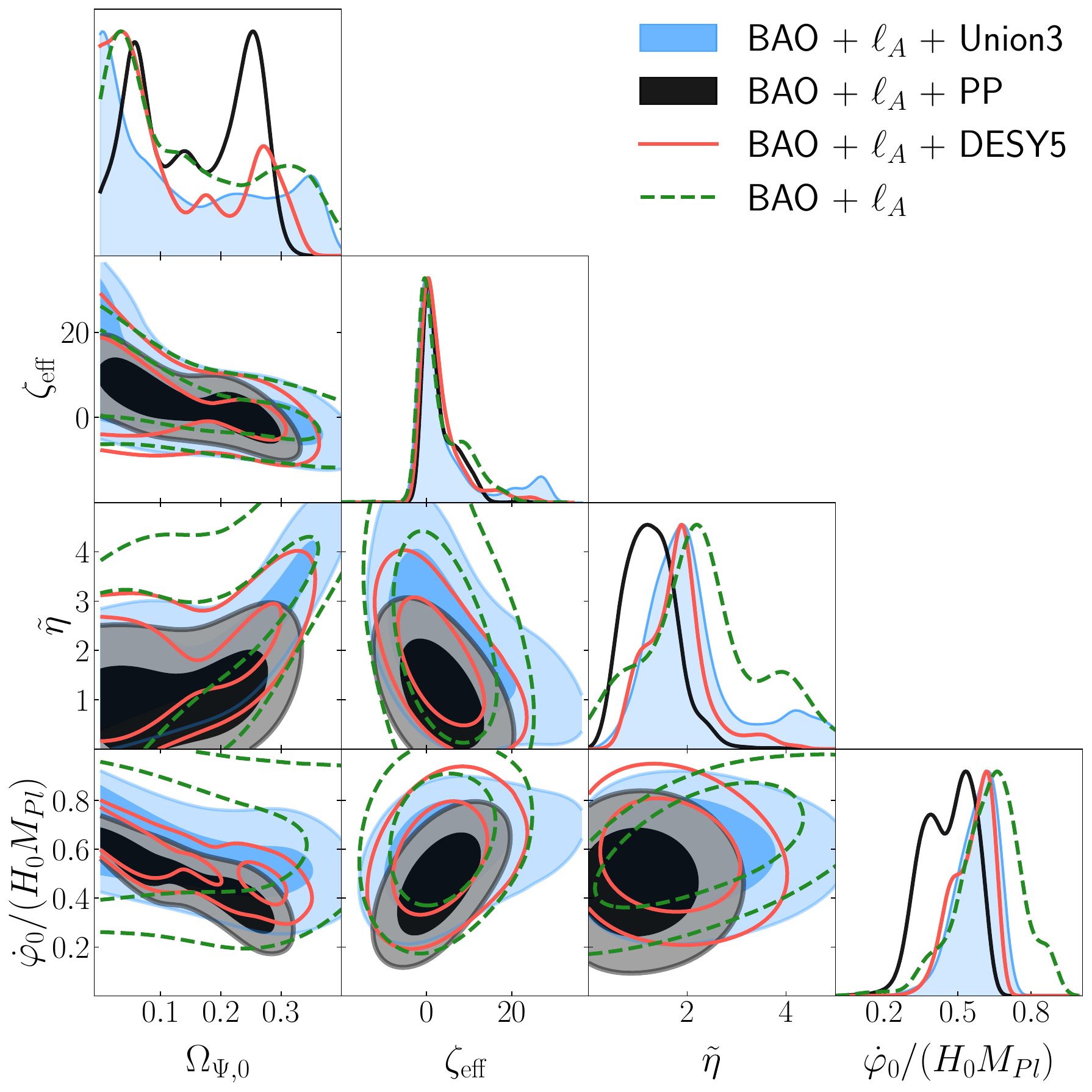}
    \caption{Marginalized posterior distributions and 68\% and 95\% confidence level (CL) contours for the main cosmological parameters $\Omega_{\Psi,0}$, $\Omega_{b_0}$, $\zeta_{\rm eff}$, $\tilde{\eta}$, and ${\dot{\varphi_0}}/({H_0M_{\text{Pl}}})$, where $\zeta_{\rm eff} = \zeta/(1-\zeta\varphi_0/M_{\rm Pl})$ and $\tilde{\eta}=\eta M_{\rm Pl}$. The results are shown for different combinations of datasets, as indicated in the legend.}
    \label{Plot_T_OP}
\end{figure}

We present the results obtained by requiring the energy densities of the fields to be strictly positive. Since certain parameter combinations can lead to negative energy densities, this particular scenario is analyzed in detail in Appendix~\ref{apA}. The observational constraints are presented in Table~\ref{Tab1}, which summarizes the marginalized mean values for the free parameters $\Omega_{\Psi,0}$, $\Omega_{b,0}$, $\zeta_{\rm eff}$, $\tilde{\eta}$, and ${\dot{\varphi_0}}/({H_0M_{\text{Pl}}})$ at the 68\% and 95\% confidence levels for different combinations of data sets. Because the posterior distributions exhibit multimodal features, the marginalized mean might not align exactly with the global maximum likelihood peak. For a complete understanding of the parameter space, these central values should be interpreted alongside the visual representations. Figure~\ref{Plot_T_OP} shows the marginalized 1$\sigma$ and 2$\sigma$ confidence contours along with the posterior distributions for the free parameters.

\begin{table*}
\centering
\small
\setlength{\tabcolsep}{3.5pt}
\renewcommand{\arraystretch}{1.6}
\begin{tabular}{lcccc}
\hline\hline
\textbf{Dataset} & \textbf{$\Omega_{\Psi,0}$} & \textbf{$\zeta_{\rm eff}$} & \textbf{$\tilde{\eta}$} & \textbf{${\dot{\varphi}_0}/({H_0M_{\text{Pl}}})$} \\
\hline
\hline
BAO + $\ell_A$ & $0.17\pm 0.13 {^{+0.23}_{-0.19}}$ & $3.3^{+5.3}_{-6.3} {^{+12}_{-7.3}}$ & $2.59^{+0.72}_{-1.8} {^{+3.1}_{-2.6}}$ & $0.62^{+0.15}_{-0.11} {^{+0.28}_{-0.28}}$ \\
BAO + $\ell_A$ + PP & $0.156\pm 0.090 {^{+0.14}_{-0.14}}$ & $2.7^{+1.4}_{-4.5} {^{+8.5}_{-4.9}}$ & $1.27^{+0.44}_{-0.57} {^{+1.0}_{-0.96}}$ & $0.46^{+0.13}_{-0.11} {^{+0.17}_{-0.18}}$ \\
BAO + $\ell_A$ + DESY5 & $0.13^{+0.17}_{-0.13} {^{+0.19}_{-0.13}}$ & $3.1^{+1.2}_{-5.1} {^{+14}_{-6.6}}$ & $1.79^{+0.46}_{-0.66} {^{+1.7}_{-1.2}}$ & $0.560^{+0.11}_{-0.088} {^{+0.13}_{-0.16}}$ \\
BAO + $\ell_A$ + Union3 & $0.16^{+0.20}_{-0.16} {^{+0.22}_{-0.17}}$ & $4.57^{+0.71}_{-8.0} {^{+25}_{-9.9}}$ & $2.18^{+0.34}_{-1.1} {^{+2.7}_{-1.6}}$ & $0.579^{+0.10}_{-0.055} {^{+0.14}_{-0.17}}$ \\
\hline\hline
\end{tabular}
\caption{Marginalized mean values and constraints on the free parameters at the $68\%$ and $95\%$ confidence levels.}
\label{Tab1}
\end{table*}

Considering the baseline BAO + $\ell_A$ dataset, we obtain $\Omega_{\Psi,0} = 0.17\pm 0.13$, at the 68\% confidence level, with the inclusion of the supernova samples, the inferred field density shifts to $\Omega_{\Psi,0} = 0.156\pm 0.090$, $\Omega_{\Psi,0} = 0.13^{+0.17}_{-0.13}$, and $\Omega_{\Psi,0} = 0.16^{+0.20}_{-0.16}$ when incorporating the PP, DESY5, and Union3 catalogs, respectively. The addition of the SNe Ia compilations leads to slightly lower central values for the field density, maintaining overall consistency. The statistical tension between the baseline BAO + $\ell_A$ outcome and the extended datasets remains minimal, corresponding to variations of 0.09$\sigma$ for PP, 0.20$\sigma$ for DESY5, and 0.05$\sigma$ for Union3. Although the central value of $\zeta_{\rm eff}$ undergoes minor fluctuations upon incorporating the SNe Ia data, all combinations demonstrate robust compatibility well within 0.2$\sigma$. A similar consistency is observed for the parameters $\tilde{\eta}$ and ${\dot{\varphi_0}}/({H_0M_{\text{Pl}}})$, where all inferred quantities across the different observational combinations agree within less than 1.0$\sigma$.

\begin{figure*}
    \centering
\includegraphics[width=0.4\textwidth]{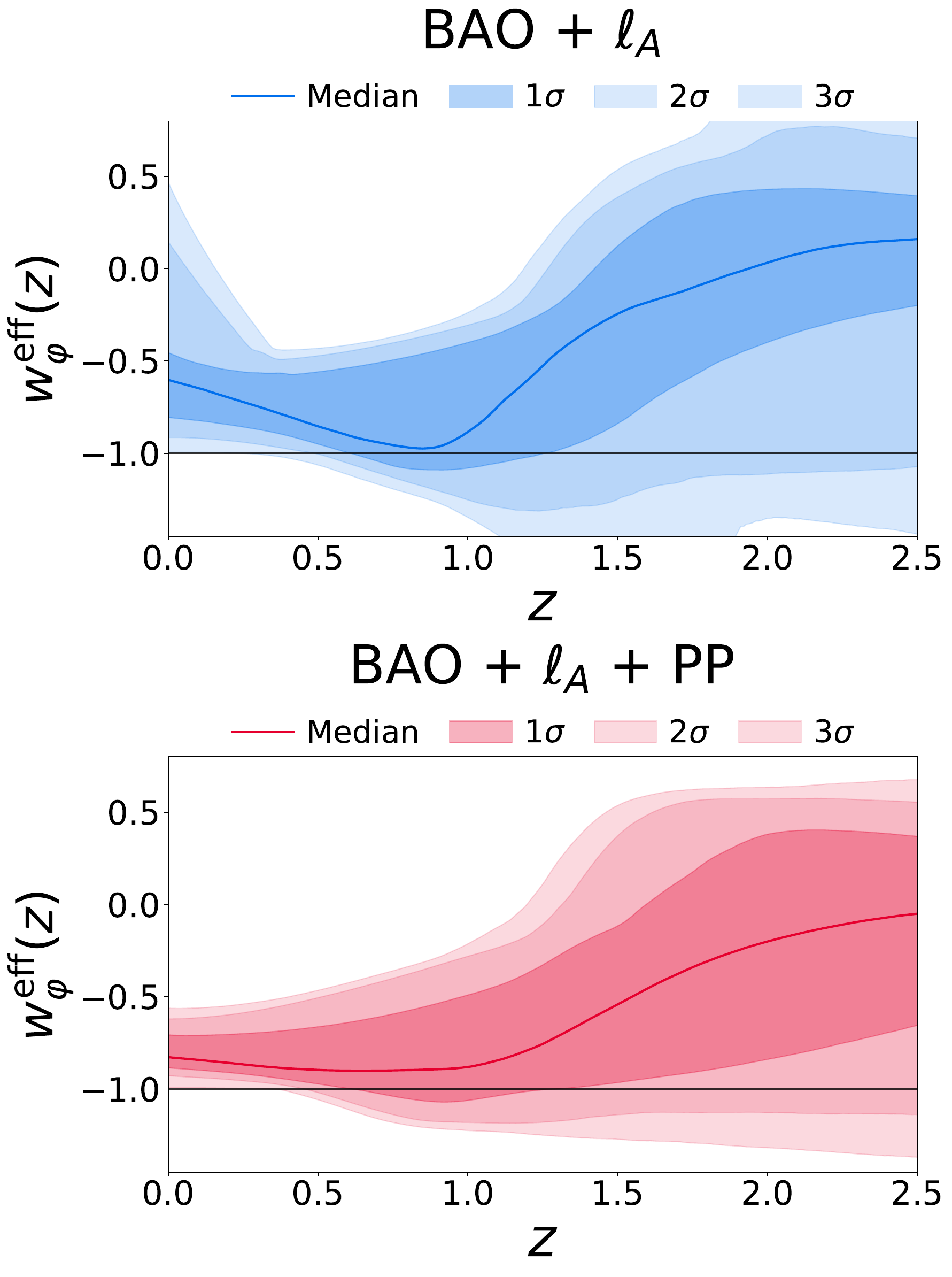}
\includegraphics[width=0.4\textwidth]{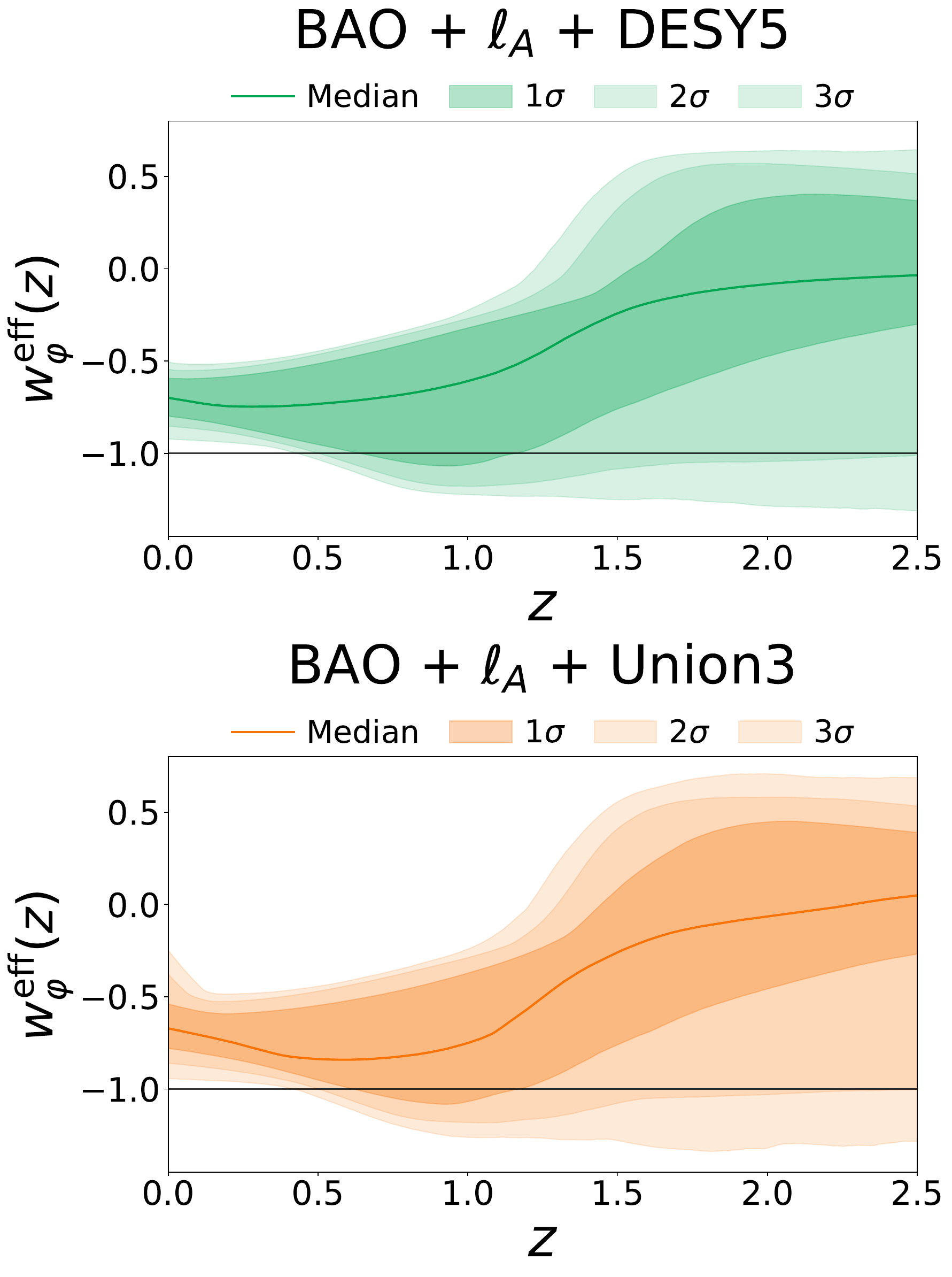} 
    \caption{Reconstruction of the evolution of $w_{\varphi}^{\text{eff}}$ from the MCMC chains, plotted up to $3\sigma$ C.I.. \textbf{Top left:} Reconstruction from BAO + $\ell_A$. \textbf{Top right:} Reconstruction from BAO + $\ell_A$ + DESY5. \textbf{Bottom left:} Reconstruction from BAO + $\ell_A$ + PP. \textbf{Bottom right:} Reconstruction from BAO + $\ell_A$ + Union3. In all panels, the solid black line represents $w = -1$.
    }
    \label{fig:weff}
\end{figure*}

We perform the reconstruction of the effective equation of state parameter $w_{\varphi}^{\mathrm{eff}}$(z), defined in Eq.~\eqref{eq:eff_EOS_Yu}, using the posterior chains obtained from our MCMC analysis. The results are displayed in Fig.~\ref{fig:weff} over the redshift interval $0<z<2.5$, with the shaded regions representing the 1$\sigma$, 2$\sigma$, and 3$\sigma$ confidence intervals around the median reconstruction, and the horizontal line at $w = - 1$ marking the phantom divide.
For the ${\rm BAO} + \ell_A$ combination (top-left panel), the 1$\sigma$ credible region crosses the phantom divide in the approximate range $0.5 \lesssim z \lesssim 1.2$, while consistency with $w = - 1$ is achieved within 2$\sigma$ over approximately 80\% of the analyzed redshift range. The ${\rm BAO} + \ell_A + {\rm PP}$ case (bottom-left panel) yields a more tightly constrained reconstruction, for which phantom crossing remains statistically permitted within 2$\sigma$ over a comparable fraction of the redshift interval.
The right panels show the reconstructions for ${\rm BAO} + \ell_A + {\rm DESY5}$ (top) and ${\rm BAO} + \ell_A + {\rm Union3}$ (bottom). Both exhibit broader uncertainty bands and display phantom divide crossing within 2$\sigma$
 across most of the redshift range, with a region of 1$\sigma$ compatibility also present at low to intermediate redshifts ($z \lesssim 0.8$). Overall, all four dataset combinations consistently indicate that, while the median reconstruction does not cross the phantom divide, a transient phantom-like phase cannot be statistically ruled out over a significant portion of the probed cosmic history.

\textit{Discussion}--A notable feature of the cosmological evolution is the rich, non-monotonic behavior of the dark energy effective EoS, which exhibits a double crossing of the phantom divide. At high redshift, the system lies in the quintessence regime, $w^{\mathrm{eff}}_\varphi>-1$, asymptotically approaching $w^{\mathrm{eff}}_\varphi\to 0$. This behavior suggests an effective matter-like behavior of dark energy at early times, arising from the interaction with fermionic dark matter.

At low redshifts, the effective equation of state can cross into the phantom regime, $w^{\mathrm{eff}}_{\varphi}<-1$. As the Universe expands, the system allows a subsequent transition back to the quintessence regime near the present epoch. As shown in Fig \ref{fig:weff}, the constraints from the Bayesian analysis indicate that such a phantom crossing is permitted within the $1\sigma$ confidence region, and is therefore statistically consistent with the data. Remarkably, this evolution is consistent with recent model-independent reconstructions and parametrized analyses from the DESI collaboration \cite{DESI:2025zgx,DESI2025DEpar}.

The condition for crossing the phantom divide follows directly from Eq.~\eqref{eq:effective_EoS}. From this equation, we see that the sign of the interaction contribution is controlled by the product $\beta\dot{\varphi}$, for $H>0$, $\rho_\varphi>0$, and $n_\Psi>0$. Therefore, the interaction can drive the effective EoS below $-1$ only when $\beta$ and $\dot{\varphi}$ have the same sign, so that $\beta\dot{\varphi}>0$. The general crossing condition of the phantom regime is
\begin{equation}
\beta\dot{\varphi} n_\Psi
>3H\alpha\dot{\varphi}^{2}\rho_\varphi .
\label{eq:phantom_condition}
\end{equation}
For $\beta\dot{\varphi}>0$, this condition becomes
\begin{equation}
|\beta| n_\Psi
>
3H\alpha|\dot{\varphi}|\rho_\varphi .
\end{equation}
By contrast, if $\beta\dot{\varphi}<0$, the interaction contribution is positive and pushes the effective EoS away from the phantom regime. The transition back to the quintessence regime occurs when the inequality in Eq.~(\ref{eq:phantom_condition}) is no longer satisfied.

Phantom behavior, therefore, emerges when the microscopic energy transfer from the fermionic sector to the scalar field overcomes both Hubble dilution and the intrinsic BI dynamics of the tachyonic field. During the phantom phase, the tachyon energy density increases due to continuous energy injection from the fermionic component. However, as the Universe expands, the fermion number density dilutes, and the interaction term responsible for the effective phantom behavior progressively weakens. The system then crosses the phantom divide back to a non-phantom regime. Hence, the double crossing is entirely dynamical and depends on the evolving balance between interaction, dilution, and the intrinsic scalar-field dynamics.

In the absence of coupling, the fermionic dark matter component still affects the expansion rate through $E(z)$, but there is no direct energy transfer between the two dark-sector fields. Consequently, the tachyonic dark energy remains confined to the quintessence regime. Thus, the double-crossing behavior observed in the coupled scenario corresponds to a transient effective phantom phase that is a purely emergent phenomenon.

Under the assumption that $\widetilde{M}$ is of order unity—a standard naturalness choice— the model expects a characteristic mass scale for the fermionic component of \begin{equation}
M \sim \sqrt{M_H M_{\rm Pl}} \simeq 1.9\times10^{-3}\,\mathrm{eV}\quad .
\end{equation}
This places the dark fermion in the ultralight regime and suggests a connection to light hidden-sector particles or sterile-neutrino-like physics. We highlight that this is not measurable only with distance data, such as the one we used in this letter. Investigations of whether this may be measured with perturbation data are outside the scope of this letter.

\textit{Conclusion}--We present a field-theoretic model of an interacting dark sector in which a Yukawa coupling between fermionic dark matter and a BI tachyonic dark energy dynamically generates an effective double crossing of the phantom divide, without introducing a fundamental phantom degree of freedom. When confronted with BAO data, CMB distance priors, and Type Ia supernovae, the posterior reconstruction of the effective dark-energy equation of state exhibits a recent phantom crossing within the $1\sigma$ standard deviation, consistent with recent DESI-motivated evidence for dynamical dark energy. The model also points to an ultralight mass scale for fermionic dark matter, which requires extending the analysis to include perturbations.

\textit{Acknowledgments}--- OPFP acknowledges financial support from Fundação de Apoio à Pesquisa do Estado da Paraíba (FAPESQ).  AAE, AM, and FBA acknowledge the support from the University of Science and Technology of China. FBA also acknowledges support from the Chinese Academy of Sciences and the Br-A Talent Program.

\appendix
\section{Unconstrained Analysis\label{apA}}

In this appendix, we evaluate the behavior of the model under a relaxed theoretical prior, specifically allowing the energy densities to cross into negative regimes. The marginalized posterior estimates for this broader parameter space are detailed in Table \ref{Tab2}. Considering the foundational combination of the BAO data and the $\ell_A$ prior, we find $\Omega_{\Psi,0} = 0.21^{+0.15}_{-0.10}$,  at the 68\% confidence level. When the different Type Ia Supernovae likelihoods are integrated into the analysis, the scalar field density experiences a slight upward trend, settling at $\Omega_{\Psi,0} = 0.248^{+0.079}_{-0.059}$ with the Pantheon+ compilation, $0.25\pm 0.11$ with DESY5, and $0.23^{+0.16}_{-0.22}$ with Union3. Despite these minor shifts in the central values, the updated constraints present no meaningful statistical disparity from the baseline results, exhibiting deviations strictly below 0.3$\sigma$. As for the effective interaction coupling $\zeta_{\rm eff}$, the posteriors favor negative central values with wide confidence margins, maintaining an agreement well within 0.45$\sigma$ across all SNe Ia additions. A comparable level of robustness is evident for the remaining variables, $\tilde{\eta}$ and ${\dot{\varphi}_0}/({H_0M_{\text{Pl}}})$. For these quantities, the largest discrepancies compared to the BAO + $\ell_A$ benchmark remain below the 0.6$\sigma$ threshold, confirming that the overall statistical consistency of the model is fully preserved even when the strict positivity bounds on the densities are lifted.
\begin{table*}
\centering
\small
\setlength{\tabcolsep}{3.5pt}
\renewcommand{\arraystretch}{1.6}
\begin{tabular}{lcccc}
\hline\hline
\textbf{Dataset} & \textbf{$\Omega_{\Psi,0}$} & \textbf{$\zeta_{\rm eff}$} & \textbf{$\tilde{\eta}$} & \textbf{${\dot{\varphi}_0}/({H_0M_{\text{Pl}}})$} \\
\hline
\hline
BAO + $\ell_A$ & $0.21^{+0.15}_{-0.10}$ & $-16^{+18}_{-23}$ & $3.73^{+0.92}_{-0.42}$ & $0.32^{+0.37}_{-0.26}$ \\
BAO + $\ell_A$ + PP & $0.248^{+0.079}_{-0.059}$ & $-19.0^{+13}_{-6.1}$ & $4.16^{+0.28}_{-0.58}$ & $0.19^{+0.20}_{-0.13}$ \\
BAO + $\ell_A$ + DESY5 & $0.25\pm 0.11$ & $-7.75^{+9.2}_{-0.90}$ & $3.7^{+2.6}_{-2.3}$ & $0.446^{+0.18}_{-0.084}$ \\
BAO + $\ell_A$ + Union3 & $0.23^{+0.16}_{-0.22}$ & $-7.13^{+9.5}_{-0.72}$ & $3.40^{+0.96}_{-1.4}$ & $0.496^{+0.20}_{-0.063}$ \\
\hline\hline
\end{tabular}
\caption{Constraints on free parameters with $68\%$ confidence limits.}
\label{Tab2}
\end{table*}

\begin{figure}
    \centering        \includegraphics[width=0.5\textwidth]{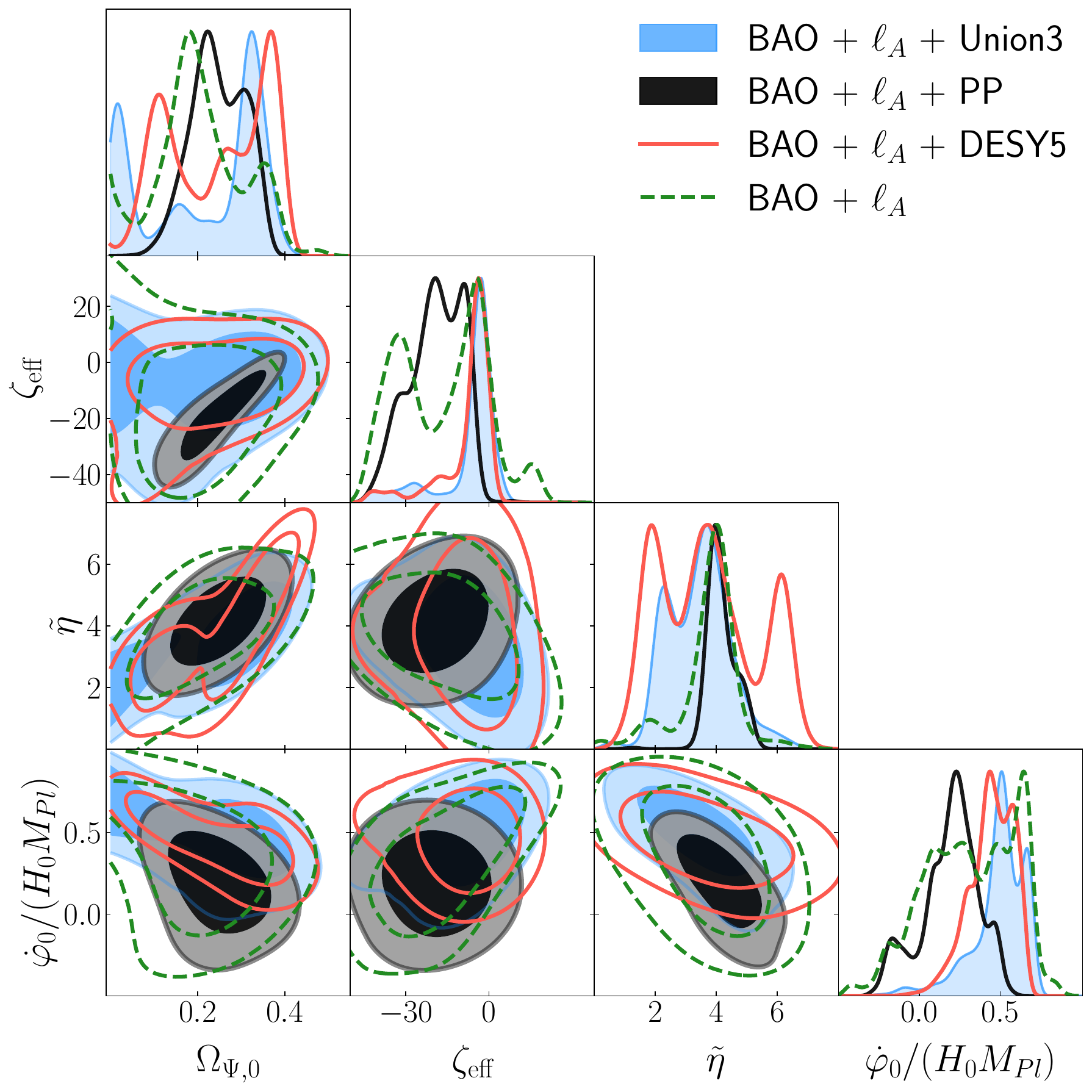} 
    \caption{Marginalized posterior distributions and 68\% and 95\% confidence level (CL) contours for the main cosmological parameters $\Omega_{\Psi,0}$, $\Omega_{b_0}$, $\zeta_{\rm eff}$, $\tilde{\eta}$, and ${\dot{\varphi_0}}/({H_0M_{\text{Pl}}})$, for the case with $\Omega_i < 0$. The results are shown for different combinations of datasets, as indicated in the legend. }
    \label{Plot_T_ON}
\end{figure}

\begin{figure*}
    \centering
\includegraphics[width=0.4\textwidth]{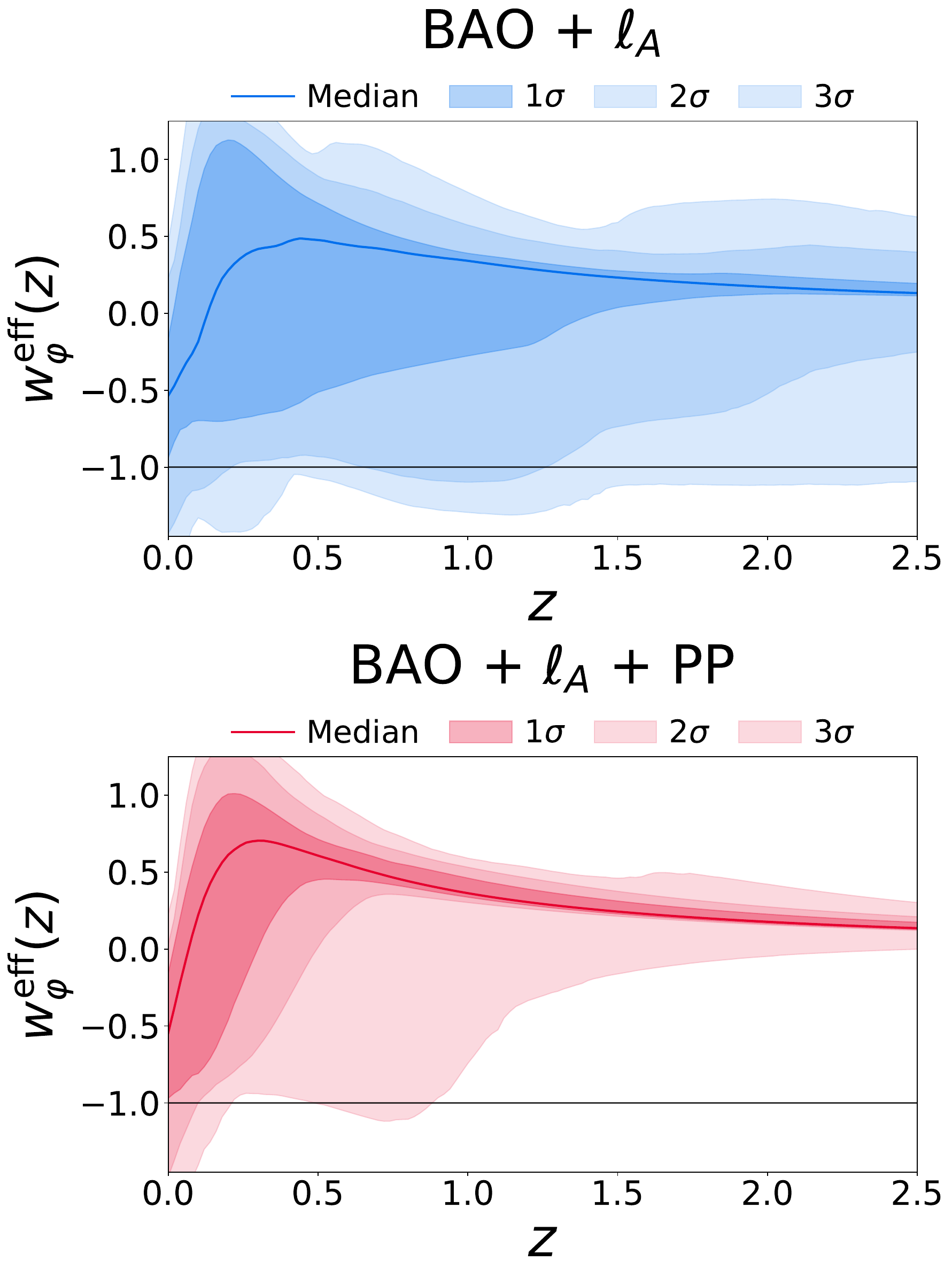}
\includegraphics[width=0.4\textwidth]{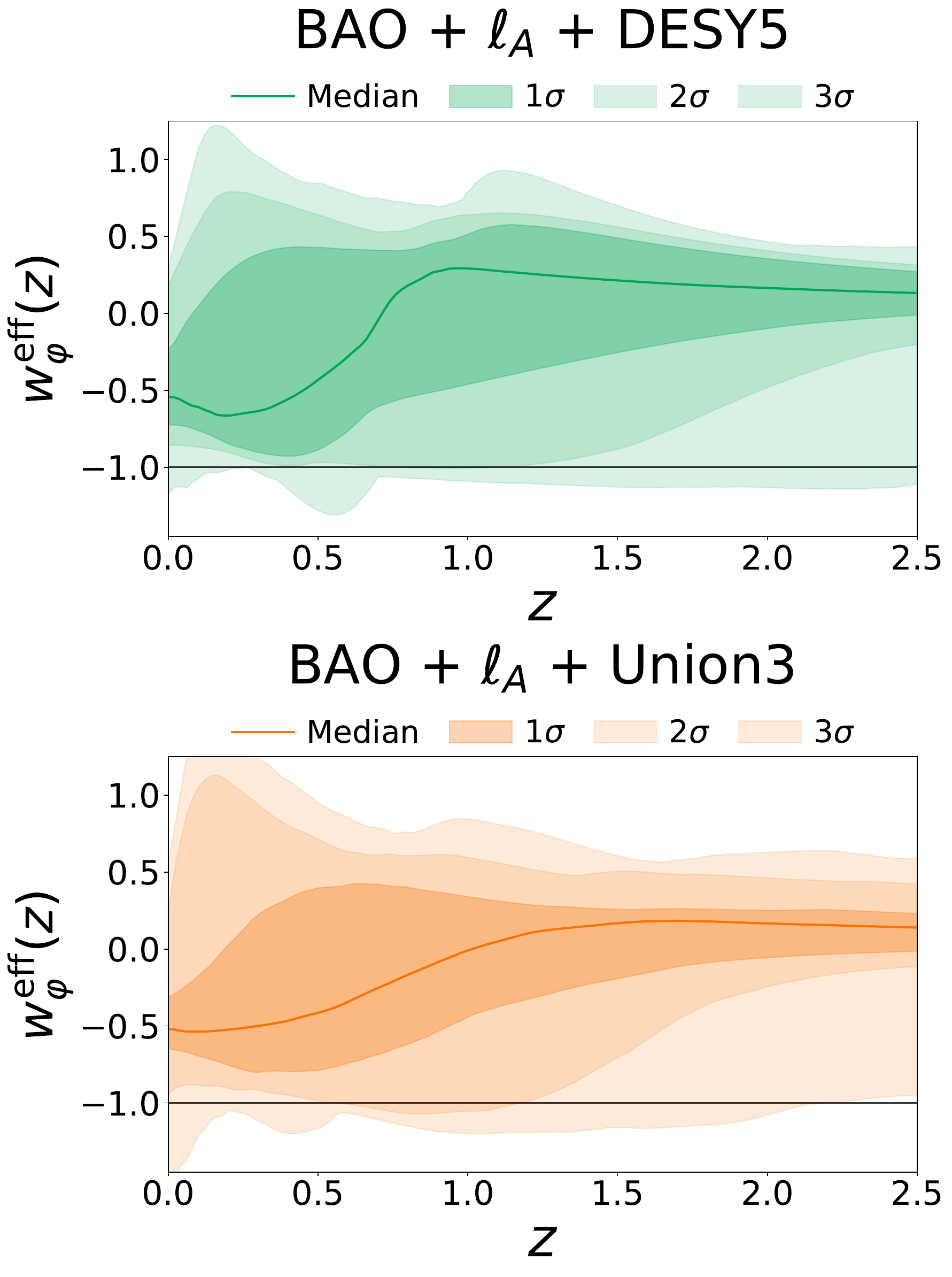} 
    \caption{Reconstruction of the evolution of $w_{\varphi}^{\text{eff}}$, for the case with $\Omega_i < 0$, from the MCMC chains, plotted up to $3\sigma$ C.I.. \textbf{Top left:} Reconstruction from BAO + $\ell_A$. \textbf{Top right:} Reconstruction from BAO + $\ell_A$ + DESY5. \textbf{Bottom left:} Reconstruction from BAO + $\ell_A$ + PP. \textbf{Bottom right:} Reconstruction from BAO + $\ell_A$ + Union3. In all panels, the solid black line represents $w = -1$.
    }
    \label{fig:weff_ON}
\end{figure*}

As shown in Figure~\ref{fig:weff_ON}, by relaxing the positivity condition on the energy densities, the reconstructed evolution of $w_{\varphi}^{\rm eff}(z)$ exhibits a distinct behavior compared to the strictly constrained scenario. The regions up to the 3$\sigma$ confidence level now accommodate significantly larger positive values for $w_{\varphi}^{\rm eff}$ across all data combinations. Furthermore, the likelihood of a phantom crossing is substantially reduced. A crossing of the phantom divide ($w = -1$) is no longer favored within the 1$\sigma$ contour at any redshift; instead, this possibility is primarily confined to the broader 2$\sigma$ and 3$\sigma$ confidence intervals in all analyzed cases.

\bibliography{apssamp}

\end{document}